\documentclass{article}
\usepackage{spconf,amsmath,graphicx}
\usepackage{epstopdf}
\usepackage{epsfig}
\usepackage{amsmath}
\usepackage{amssymb}
\usepackage{bm}
\usepackage{subfigure}
\usepackage{url}
\usepackage{color}

\title{Learned Lossless Image Compression with a HyperPrior and Discretized Gaussian Mixture Likelihoods}
\name{Zhengxue Cheng, Heming Sun, Masaru Takeuchi, Jiro Katto}\address{Department of Computer Science and Communications Engineering, Waseda University, Tokyo, Japan.}

\begin{document}
%
\maketitle
\begin{abstract}
Lossless image compression is an important task in the field of multimedia communication. Traditional image codecs typically support lossless mode, such as WebP, JPEG2000, FLIF. Recently, deep learning based approaches have started to show the potential at this point. HyperPrior is an effective technique proposed for lossy image compression. This paper generalizes the hyperprior from lossy model to lossless compression, and proposes a L2-norm term into the loss function to speed up training procedure. Besides, this paper also investigated different parameterized models for latent codes, and propose to use Gaussian mixture likelihoods to achieve adaptive and flexible context models. Experimental results validate our method can outperform existing deep learning based lossless compression, and outperform the JPEG2000 and WebP for JPG images.
\end{abstract}

\vspace{-2mm}
\section{Introduction}
\label{sec:intro}

Image compression is a fundamental task in the field of signal processing in many decades for efficient transmission and storage. Classical image compression standards, such as JPEG~\cite{IEEEexample:JPEG}, JPEG2000~\cite{IEEEexample:JPEG2000}, WebP~\cite{IEEEexample:webp}, HEVC/H.265-intra~\cite{IEEEexample:HEVC} and lossless FLIF~\cite{IEEEexample:flif}, usually rely on hand-crafted encoder/decoder (codec) block diagrams. They use fixed linear transform, quantization and the entropy coder to reduce spatial redundancy for images. Typically they also support lossless model to compress images lossless. However, along with the fast development of new image formats and high-resolution mobile devices, existing image compression algorithms are not expected to be optimal solutions.

Recently, we have seen a great surge of deep learning based image compression approaches. Most of them focus on lossy image compression. For example, some image compression approaches use generative models to learn the distribution of images using adversarial training~\cite{IEEEexample:waveone, IEEEexample:MITgan, IEEEexample:Extreme} to achieved impressive subjective quality at extremely low bit rate. Some works use recurrent neural networks to compress the residual information recursively, such as~\cite{IEEEexample:Toderici01, IEEEexample:Toderici, IEEEexample:Nick} to realize scalable coding. Some approaches propose a hyperprior-based and context-adaptive context model to compress codes effectively in~\cite{IEEEexample:Balle2,IEEEexample:David,IEEEexample:Lee}. Some methods decorrelate each channel of latent codes and apply deep residual learning to improve the performance as~\cite{IEEEexample:PCS,IEEEexample:ourCVPR,IEEEexample:ourCLIC}. However, deep learning based lossless compression has rarely discussed. One related work is L3C~\cite{IEEEexample:l3c} to propose a hierarchical architecture with $3$ scales to compress images lossless.

In this paper, we propose a learned lossless image compression using a hyperprior and discretized Gaussian mixture likelihoods. Our contributions mainly consist of two aspects. First, we generalize the hyperprior from lossy model to lossless compression model, and propose a loss function with L2-norm for lossless compression to speed up training. Second, we investigate four parameterized distributions and propose to use Gaussian mixture likelihoods for the context model. Experimental results have demonstrated our method can outperform recent learned compression approach L3C. Besides, our method outperform JPEG2000 and WebP for JPG images.


\vspace{-2mm}
\section{Learned Lossless Image Compression }
\vspace{-2mm}
\subsection{Formulation of Compression with a Hyperprior}

\begin{figure*}[tb]
\centering
\subfigure[Lossy model]{
\label{Fig.op.1}
\includegraphics[width=0.18\textwidth]{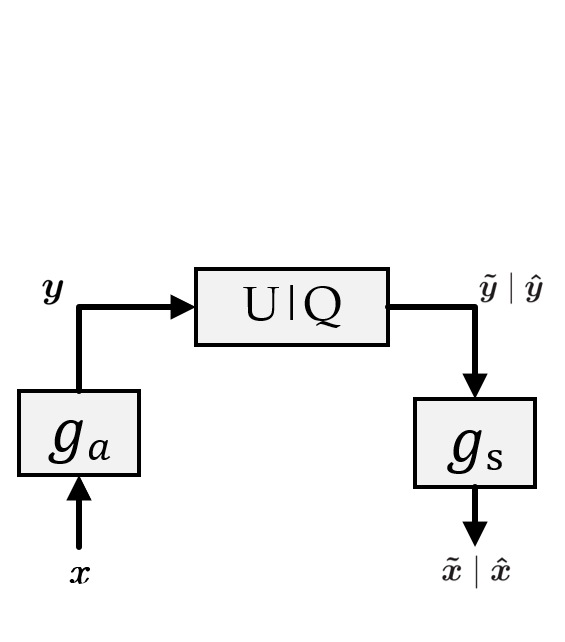}}
\hspace{0.15in}
\subfigure[Lossy model with a hyperprior]{
\label{Fig.op.2}
\includegraphics[width=0.20\textwidth]{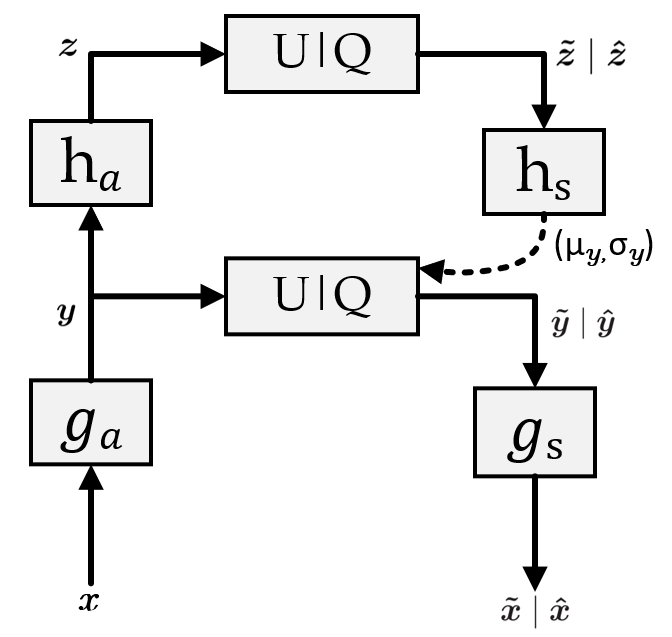}}
\hspace{0.15in}
\subfigure[Proposed Lossless model with a hyperprior]{
\label{Fig.op.3}
\includegraphics[width=0.20\textwidth]{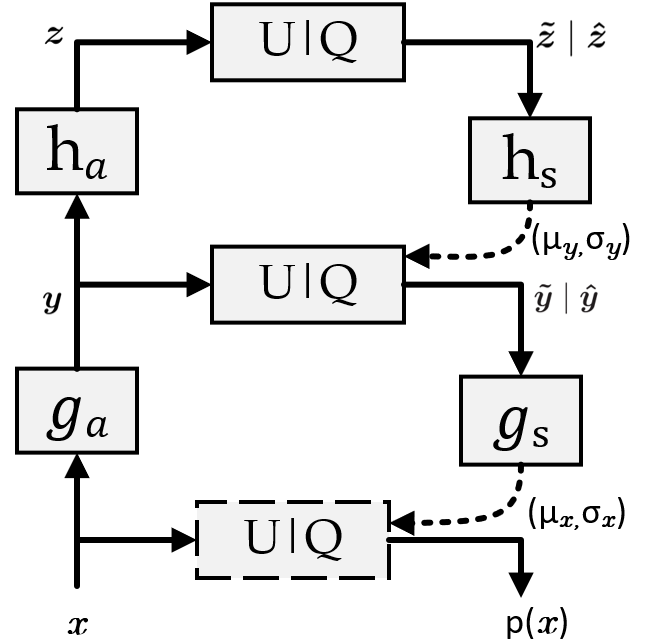}}
\hspace{0.15in}
\subfigure[Distributions of lossy and lossless models]{
\label{Fig.op.4}
\includegraphics[width=0.18\textwidth]{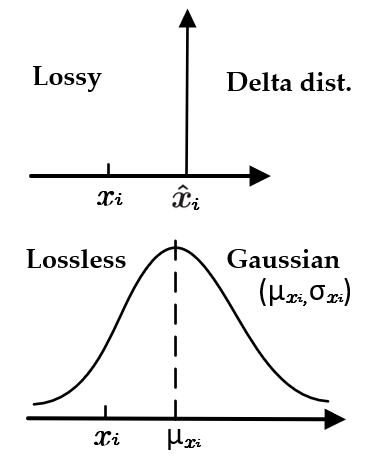}}
\caption{Operational diagrams and distributions of learned lossy and lossless image compression.}
\label{fig:op}
\end{figure*}


According to the transform coding approach~\cite{IEEEexample:transformcoding}, image compression can be formulated as
\begin{equation}
\begin{aligned}
\boldsymbol{y} &=g_{a}(\boldsymbol{x}; \boldsymbol{\phi}) \\
\hat{\boldsymbol{y}} &= Q(\boldsymbol{y}) \\
\hat{\boldsymbol{x}} &=g_{s}(\hat{\boldsymbol{y}}; \boldsymbol{\theta})
\end{aligned}
\end{equation}
where $\boldsymbol{x}$, $\hat{\boldsymbol{x}}$, $\boldsymbol{y}$, and $\hat{\boldsymbol{y}}$ are raw images, reconstructed images, a latent presentation before quantization, and compressed codes, respectively. $Q$ represents the quantization , which is approximated by a uniform noise $\mathcal{U}(-\frac{1}{2}, \frac{1}{2})$ during the training, which is denoted as $U|Q$ in Fig.~\ref{fig:op}. $\boldsymbol{\phi}$ and $\boldsymbol{\theta}$ are optimized parameters of analysis and synthesis transforms. The operation diagram is explained in Fig.~\ref{Fig.op.1}.

In the work~\cite{IEEEexample:Balle2}, \emph{Ball\'e} proposed a hyperprior for lossy image compression to achieve promising results. Hyperprior introduces a side information $\boldsymbol{z}$ to capture spatial dependencies among the elements of $\boldsymbol{y}$, formulated as
\begin{equation}
\begin{aligned}
&\boldsymbol{z}=h_{a}(\boldsymbol{y}; \boldsymbol{\phi_{h}}); \;\;\\
&\hat{\boldsymbol{z}} = Q(\boldsymbol{z}) \\
(&\boldsymbol{\mu_y}, \boldsymbol{\sigma_y}) =h_{s}(\hat{\boldsymbol{z}}; \boldsymbol{\theta_{h}})\\
\end{aligned}
\end{equation}
where $h_{a}$ and $h_{s}$ denote the analysis and synthesis functions in the auxiliary autoencoder, where $\boldsymbol{\phi_{h}}$ and $\boldsymbol{\theta_{h}}$ are optimized parameters of the hyperprior, respectively. $\boldsymbol{\mu_y}$ and $\boldsymbol{\sigma_y}$ are generated by the auxiliary autoencoder to model a Gaussian distribution $\mathcal{N}(\boldsymbol{\mu_y}, \boldsymbol{\sigma_y}^{2})$. Originally, \emph{Ball\'e} assumed a zero-mean scalar Gaussian distribution, but in the enhanced work~\cite{IEEEexample:David}, a mean and scale Gaussian distribution achieves better results. Then we use a mean and scale one in this paper. The operation diagram is explained in Fig.~\ref{Fig.op.2}.

However, both cases are lossy compression. In this paper, we generalize a lossy model to a lossless model as Fig.~\ref{Fig.op.3}. Not only the reconstructed pixel value, we predict a probability model for raw images, and model $\boldsymbol{x}$ using another parameterized distributions. This is feasible because entropy coding techniques such as arithmetic coding~\cite{IEEEexample:arithmeticcoding} can losslessly compress the signals if a probability model is given. The distribution difference of lossy and lossless compression is illustrated in Fig.~\ref{Fig.op.4}, where we can think lossy compression is to predict a delta distribution at the value $\hat{\boldsymbol{x}}$ for each element, while lossless compression is to predict a more generalized and arbitrary probability models. One intuitive choice is to use Gaussian distribution $\mathcal{N}(\boldsymbol{\mu_x}, \boldsymbol{\sigma_x}^{2})$, like $\boldsymbol{y}$, as Fig.~\ref{Fig.op.4}.

\begin{figure*}[tb]
	\centerline{\psfig{figure=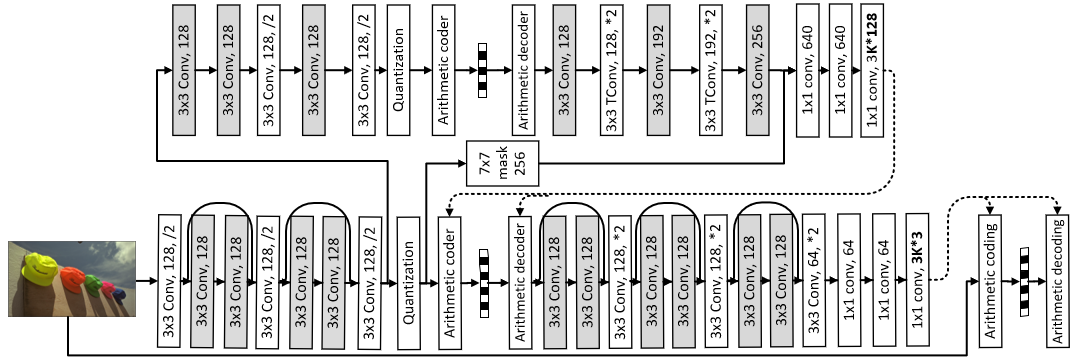,width=162mm} }
	\caption{Network architecture of lossless image compression.}
	\label{fig:net}
\end{figure*}

\vspace{-2mm}
\subsection{Proposed Loss Function with a L2-Norm}

Different from lossy compression which controls the rate-distortion tradeoff, the loss function of lossless compression only consists of required bits for $\boldsymbol{x}$, $\boldsymbol{y}$ and $\boldsymbol{z}$, that is,
\begin{equation}
\begin{aligned}
\mathcal{L} = &\mathop{\mathbb{E}}[-\log_{2}(p_{\hat{\boldsymbol{x}}}(\hat{\boldsymbol{x}}|\boldsymbol{\mu_{x}},\boldsymbol{\sigma_{x}} ))] + \mathop{\mathbb{E}}[-\log_{2}(p_{\hat{\boldsymbol{y}}}(\hat{\boldsymbol{y}}|\boldsymbol{\mu_{y}},\boldsymbol{\sigma_{y}} ))] + \\ &\mathop{\mathbb{E}}[-\log_{2}(p_{\hat{\boldsymbol{z}}}(\hat{\boldsymbol{z}}|\boldsymbol{\psi} ))]
\end{aligned}
\end{equation}
where the probability models of $\boldsymbol{\hat{x}}$ are estimated by the parameters $\boldsymbol{\mu_{x}},\boldsymbol{\sigma_{x}}$, which are conditioned on $\boldsymbol{\hat{y}}$ actually. Similarly, the probability model of $\boldsymbol{y}$ are estimated by parameters conditioned on $\boldsymbol{\hat{z}}$, i.e.
\begin{equation}
\begin{aligned}
&p_{\hat{\boldsymbol{x}}}(\hat{\boldsymbol{x}}|\hat{\boldsymbol{y}}) = \prod_{i} (\mathcal{N}(\mu_{x_i}, \sigma_{x_i}^{2})\ast \mathcal{U}(-\frac{1}{2}, \frac{1}{2}))(\hat{x}_{i}) \\
\end{aligned}
\end{equation}
\begin{equation}
\begin{aligned}
&p_{\hat{\boldsymbol{y}}}(\hat{\boldsymbol{y}}|\hat{\boldsymbol{z}}) = \prod_{i} (\mathcal{N}(\mu_{y_i}, \sigma_{y_i}^{2})\ast \mathcal{U}(-\frac{1}{2}, \frac{1}{2}))(\hat{y}_{i}) \\
\end{aligned}
\end{equation}
\begin{equation}
\begin{aligned}
&p_{\hat{\boldsymbol{z}}}(\hat{\boldsymbol{z}}|\boldsymbol{\psi}) = \prod_{i} (p_{z_{i}|\boldsymbol{\psi}_{i}}(\boldsymbol{\psi}_{i})\ast \mathcal{U}(-\frac{1}{2}, \frac{1}{2}))(\hat{z}_{i}) \\
\end{aligned}
\end{equation}
where $\boldsymbol{\psi}$ are learnable factorized distributions~\cite{IEEEexample:Balle2} for $\boldsymbol{z}$ because there is no prior for $\boldsymbol{z}$.

Basically, by using Eq.(3), we can train a neural network for lossless compression. However, by experiments, we have found the converge is very slow. By observing the distribution in the distribution of Fig.~\ref{Fig.op.4}, we can notice the actual marginal distribution of $\boldsymbol{x}$ typically is not identical to the estimated entropy model $p_{\hat{\boldsymbol{x}}}$, especially for the initial training stage of neural networks. It takes quite a long time to find a proper and accurate distribution, which is centralized at the actual $\boldsymbol{x}$ value. Therefore, we novelly introduce a L2-norm term between the ground-truth value and predicted mean value into the lossless compression, that is,
\begin{equation}
\mathcal{L'} =  ||\mu_{\boldsymbol{x}} - \boldsymbol{\hat{x}}||^{2} + ||\mu_{\boldsymbol{y}} - \boldsymbol{\hat{y}}||^{2}
\end{equation}
Then the updated loss function for lossless image compression is defined as
\begin{equation}
\begin{aligned}
\label{eq.loss}
\mathcal{L} = &\mathop{\mathbb{E}}[-\log_{2}(p_{\hat{\boldsymbol{x}}}(\hat{\boldsymbol{x}}|\hat{\boldsymbol{y}})) + \mathop{\mathbb{E}}[-\log_{2}(p_{\hat{\boldsymbol{y}}}(\hat{\boldsymbol{y}}|\hat{\boldsymbol{z}} )) + \\ &\mathop{\mathbb{E}}[-\log_{2}(p_{\hat{\boldsymbol{z}}}(\hat{\boldsymbol{z}}|\boldsymbol{\psi} )) + \lambda\cdot(||\mu_{\boldsymbol{x}} - \boldsymbol{\hat{x}}||^{2} + ||\mu_{\boldsymbol{y}} - \boldsymbol{\hat{y}}||^{2})
\end{aligned}
\end{equation}
where $\lambda$ controls the weights of L2-norm term.

\subsection{Proposed Discretized Gaussian Mixture Likelihoods}

\begin{figure}[tb]
\centering
\subfigure[sub-pixel values $\boldsymbol{x}$]{
\label{Fig.x.1}
\includegraphics[height=3.1cm]{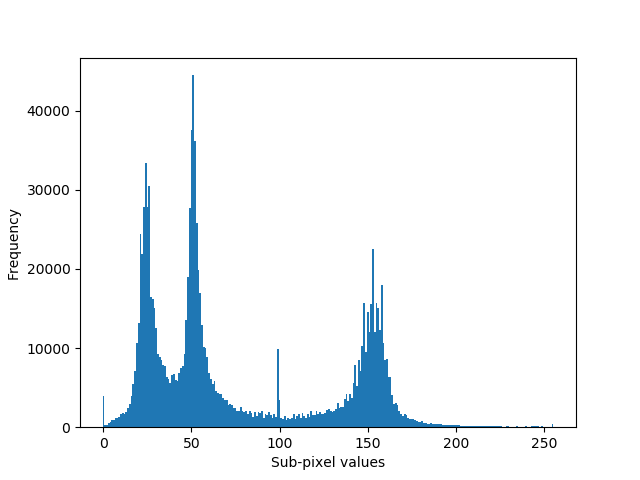}}
\subfigure[compressed codes $\boldsymbol{\hat{y}}$]{
\label{Fig.y.2}
\includegraphics[height=3.1cm]{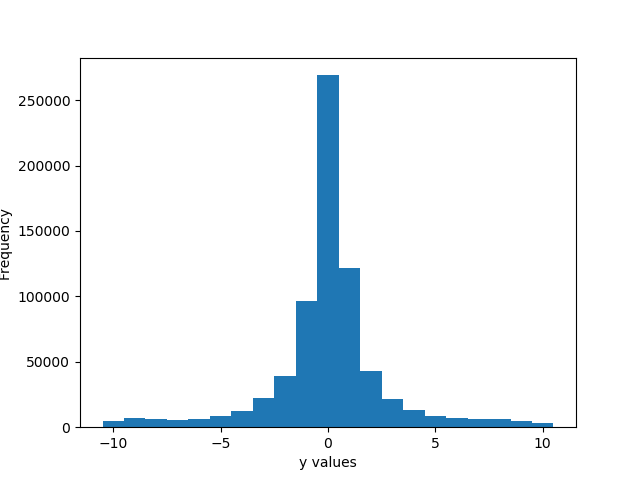}}
\caption{Margin Distributions for Image \emph{kodim02}.}
\label{fig:margin}
\end{figure}

Whether the parameterized distribution model fits the margin distribution of $\boldsymbol{x}$ and $\hat{\boldsymbol{y}}$ is a key factor for the performance. We visualize the margin distribution of all sub-pixel values $\boldsymbol{x}$ and compressed codes $\boldsymbol{\hat{y}}$ for \emph{kodim02} from Kodak dataset~\cite{IEEEexample:kodak} in Fig.~\ref{fig:margin}, and it does not actually satisfy the assumption that they follow Gaussian distribution. Previous works have investigated several parameterized distribution models, but to our knowledge, which distribution model can fit true margin distribution best is still an open question. For example, standard PixelCNN~\cite{IEEEexample:pixelcnn} uses full 256-way softmax likelihoods, which is memory-consuming and impractical for large images. PixelCNN++~\cite{IEEEexample:pixelcnnplus} proposed a discretized logistic mixture likelihoods to achieve faster training. L3C~\cite{IEEEexample:l3c} followed the PixelCNN++ to use a logistic mixture model. Lossy compression work~\cite{IEEEexample:Balle2} assumed a univariate Gaussian distribution. The work~\cite{IEEEexample:Tucodec} used Laplace distribution. Traditional coding work~\cite{IEEEexample:cauchy} used a Cauchy distribution to model coefficients.

\begin{figure}[tb]
	\centerline{\psfig{figure=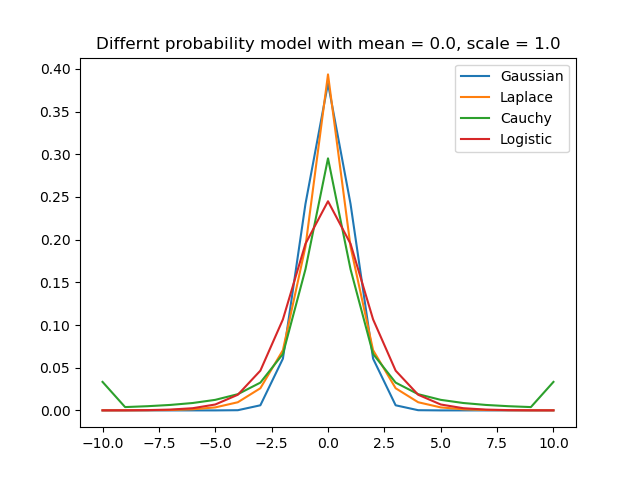,width=55mm} }
	\caption{Different distributions with zero-mean and one-scale.}
	\label{fig:distributions}
\end{figure}

We investigate all the above models, including Gaussian distribution, Cauchy distribution, Laplace distribution and Logistic distribution. We replace them for $\hat{\boldsymbol{y}}$ individually. Examples of their distribution models are visualized in Fig.~\ref{fig:distributions}, where we clip the ranges and add the tail masses to edge cases. It can be observed Cauchy distribution has longer tail than other distributions. Corresponding results are listed in Table~\ref{table.distributions} and performance is evaluated in terms of bits per sub-pixel (bpsp) following the L3C on Kodak dataset. Gaussian distribution can achieve the best performance among them.

\begin{table}[h]
\centering
\caption{The effect of different distribution models}
\label{table.distributions}
\begin{tabular}{ccccc}
\hline
Model & \textbf{\emph{Gaussian}} & \emph{Cauchy} & \emph{Laplace} & \emph{Logistic} \\
\hline
bpsp  & \textbf{3.568}   &4.189    &3.778  &3.814\\
\hline
\end{tabular}
\end{table}

To enhance the performance, we further improve the parameterized probabilities from univariate Gaussian model to multivariate Gaussian mixture model as
\begin{equation}
\begin{aligned}
&p_{\hat{\boldsymbol{y}}}(\hat{\boldsymbol{y}}|\hat{\boldsymbol{z}}) = \prod_{i} p_{\hat{\boldsymbol{y}}}(\hat{y}_{i}|\hat{\boldsymbol{z}})\\
&p_{\hat{\boldsymbol{y}}}(\hat{y}_{i}|\hat{\boldsymbol{z}}) = (\sum_{k=1}^{K}w_{y_i}^{(k)}\mathcal{N}(\mu_{y_i}^{(k)}, \sigma_{y_i}^{2(k)})\ast \mathcal{U}(-\frac{1}{2}, \frac{1}{2}))(\hat{y}_{i})\\
\end{aligned}
\end{equation}
where the $k$-th mixture distribution is characterized by Gaussian distribution with three parameters, i.e. weights $w_{i}^{(k)}$, means $\mu_{i}^{(k)}$ and variances $\sigma_{i}^{2(k)}$ for each $i$-th element. The mixture models for $\boldsymbol{x}$ are conducted in the same way as $\boldsymbol{y}$. Multivariate mixture model offers more flexibility than univariate one to model arbitrary likelihoods, as Fig.~\ref{Fig.x.1}.

\begin{table*}[htb]
\centering
\caption{Compression performance on CLIC Professional, CLIC Mobile and Kodak datasets.}
\label{table.bpsp2}
\begin{tabular}{p{40mm}p{26mm}p{24mm}p{24mm}p{24mm}}
\hline
[JPG] &  \textbf{Method} & \textbf{\emph{CLICP}}  & \textbf{\emph{CLICM}}   & \textbf{\emph{Kodak}}\\
\hline
Non-Learned Methods & JPEG2000 &2.936 \,{\footnotesize\color{blue}+7.7\%}  &2.897 \,{\footnotesize\color{blue}+9.0\%} & 2.822 \,{\footnotesize\color{blue}+7.0\%}  \\
\cline{2-5}
& WebP      &2.777 \,{\footnotesize\color{blue}+1.9\%}   &2.735 \,{\footnotesize\color{blue}+2.9\%} & 2.708 \,{\footnotesize\color{blue}+2.7\%}   \\
\cline{2-5}
& FLIF      &2.631 \,{\footnotesize\color{red}-3.5\%}    &2.516  \,{\footnotesize\color{red}-5.4\%}     & 2.479    \,{\footnotesize\color{red}-6.1\%}     \\
\hline
Learned Methods & L3C   &2.739 \,{\footnotesize\color{blue}+0.5\%}     & 2.647 \,{\footnotesize\color{red}-0.5\%}    & 2.730  \,{\footnotesize\color{blue}+3.5\%}  \\
\cline{2-5}
& \textbf{Proposed}    &\textbf{2.726}   &\textbf{2.659}      &\textbf{2.638}\\
\hline
\end{tabular}
\end{table*}

\vspace{-2mm}
\section{Experiments}

\subsection{Implementation details}

The network architectures are illustrated in Fig.~\ref{fig:net}. This residual learning architecture refers to~\cite{IEEEexample:ourCLIC} with competitive performance. We include one mask convolution to $\boldsymbol{y}$ as~\cite{IEEEexample:David}. In our experiments, $K$ is set as $3$. We set $\lambda$ in Eq.~(\ref{eq.loss}) as $0.6$ for warm-up ($8\times10^{4}$ steps in our experiments) and set $\lambda$ as $0$ after that, because optimizing this L2-norm basically contributes to smaller $\mathcal{L}$, instead, optimizing $\mathcal{L}$ not absolutely requires the minimized L2-norm term at the late training stage. For instance, some symbols with very large variance would not put too much constraint on accurate mean estimation. This L2-norm is similar to mean square error (MSE) distortion in lossy image compression. To some degree, lossless compression is a generalized form of lossy one, because lossy compression only samples the variable $\hat{\boldsymbol{x}}$ from $p_{\hat{\boldsymbol{x}}}$ with the largest probability, while lossless compression predicts the full likelihoods.

For training, we use about $35000$ patches with the size of $128\times128$ cropped from ImageNet ILSVRC dataset~\cite{IEEEexample:ImageNet}. Batch size is $8$. The model was optimized using Adam~\cite{IEEEexample:adam} algorithm. In addition, the learning rate was fixed at $1\times10^{-4}$ and was reduced to $1\times10^{-5}$ for the last $80k$ steps. We train up to $6.8\times10^{5}$ steps to achieve stable performance.


\subsection{Performance Comparison and Discussion}

For evaluation, we use three datasets, CLIC~\cite{IEEEexample:CLIC} professional validation dataset with $41$ images, CLIC mobile validation dataset with $61$ images, and Kodak dataset~\cite{IEEEexample:kodak} with 24 images to cover many different kinds of contents. Following~\cite{IEEEexample:l3c}, we resize the high-resolution CLIC images to $768$ pixels for the longer side. We compare our method with classical compression algorithms FLIF~\cite{IEEEexample:flif}, WebP 1.0.2~\cite{IEEEexample:webp}, and OpenJPEG (JPEG2000 official software)~\cite{IEEEexample:JPEG2000}, and L3C~\cite{IEEEexample:l3c}. For each model, average bits per sub-pixel (bpsp) are measured on average across all the test images.


Currently the format of large-scale training image dataset is JPG, which has already been largely compressed with many artifacts, such as smoothing areas, blocking and ringing artifacts. Even though some artifacts are invisible, they largely affect the distributions. Therefore, to evaluate the performance more close to what we train, we decide to evaluate all these methods in two ways. First, we convert test set to JPG images with quality level as 95 using PIL library for evaluation, which followed L3C~\cite{IEEEexample:l3c} way. Second, we evaluate results with lossless PNG images to show the performance gap when the test set is much different from training set.

The performance comparison on JPG images is shown in Table~\ref{table.bpsp2}. It can be observed that our method achieves better performance than JPEG2000 and WebP for all the dataset. Our method is better than L3C for \emph{CLICP} and \emph{Kodak}, but is slightly worse than L3C for \emph{CLICM}. On the other hand, the performance on lossless images is listed in Table~\ref{table.bpsp}. Our method is still better than L3C. Although our performance is worse than JPEG2000 and WebP, it comes from the difference between training datasets and test datasets. As a future work, finding some ways to remove inherent compression artifacts of training images or build a lossless dataset is worthy trying.

\begin{table}[tb]
\centering
\caption{Compression performance on lossless images.}
\label{table.bpsp}
\begin{tabular}{lllll}
\hline
[PNG] &  \textbf{Method}  & \textbf{\emph{CLICP}}  & \textbf{\emph{CLICM}} & \textbf{\emph{Kodak}} \\
\hline
Non-Learned  & PNG &4.298   &4.374   & 4.350     \\
\cline{2-5}
& JPEG2000 &3.403  &3.266   & 3.191 \\
\cline{2-5}
& WebP &3.254   &3.212  & 3.206  \\
\cline{2-5}
& FLIF &3.141   & 3.083  & 2.903   \\
\hline
Learned  & L3C  &4.098    & 3.691   & 3.547   \\
\cline{2-5}
& \textbf{Proposed}  &\textbf{3.647} &\textbf{3.486} &\textbf{3.475}   \\
\hline
\end{tabular}
\end{table}

%
%
%

\subsection{Relation to Prior Work}

The most related works are hyperprior~\cite{IEEEexample:Balle2} and L3C~\cite{IEEEexample:l3c}. Compared to~\cite{IEEEexample:Balle2}, we generalize the hyperprior from a lossy compression model to a lossless one. L3C is a recent learned lossless compression with $3$ scales, while our architecture is somewhat like a simplified hierarchical network with $2$ scales (i.e. $\boldsymbol{y}$ and $\boldsymbol{z}$). Except for the difference on architectures, more importantly, we propose two novel strategies to improve the performance. 1) We add a L2-norm to loss function which contributes to stable and fast training. 2) We used Gaussian mixture model, while L3C used Logistic mixture model. We have discussed $4$ parameterized models in Section 2.3 to show Gaussian achieves slightly better performance.

\section{Conclusion}

In this paper, we propose a learned lossless image compression using a hyperprior and discretized Gaussian mixture likelihoods. First, we formulate a hyperprior-based lossless image compression model. Based on it, we propose corresponding loss function with L2-norm to speed up training. Second, the determination of parameterized distributions is a significant factor for performance. We investigate 4 types and propose to use Gaussian mixture likelihoods. Experimental results have demonstrated our method can outperform recent learned compression approach L3C. Besides, our method outperform JPEG2000 and WebP for JPG images.

\section{Acknowledgement}

The authors would like to thank Fabian Mentzer (the first author of L3C) for fruitful discussion and insightful feedback on the evaluation methods and datasets.

\vfill\pagebreak

%

\bibliographystyle{IEEEbib}

\end{document}